\documentclass[proof]{WileyASNA-v1}

\usepackage{tikz, graphicx}	
\usepackage{amsmath}	
\usepackage{hyperref}
\usepackage{tabularx}
\usepackage{subcaption}
\usepackage{wasysym}
\usepackage{siunitx}
\usepackage[
    starfontserif 
    ]{starfont}

\DeclareSymbolFont{starfontsym}{OT1}{sts}{m}{n}
\DeclareMathSymbol{\mathSun}{\mathord}{starfontsym}{115}
\DeclareMathSymbol{\mathMercury}{\mathord}{starfontsym}{102}
\DeclareMathSymbol{\mathVenus}{\mathord}{starfontsym}{103}
\DeclareMathSymbol{\mathTerra}{\mathord}{starfontsym}{76}
\DeclareMathSymbol{\mathvarTerra}{\mathord}{starfontsym}{108}
\DeclareMathSymbol{\mathMoon}{\mathord}{starfontsym}{100}
\DeclareMathSymbol{\mathvarMoon}{\mathord}{starfontsym}{97}
\DeclareMathSymbol{\mathMars}{\mathord}{starfontsym}{104}
\DeclareMathSymbol{\mathJupiter}{\mathord}{starfontsym}{106}
\DeclareMathSymbol{\mathSaturn}{\mathord}{starfontsym}{83}
\DeclareMathSymbol{\mathUranus}{\mathord}{starfontsym}{70}
\DeclareMathSymbol{\mathvarUranus}{\mathord}{starfontsym}{65}
\DeclareMathSymbol{\mathNeptune}{\mathord}{starfontsym}{71}
\DeclareMathSymbol{\mathPluto}{\mathord}{starfontsym}{74}
\DeclareMathSymbol{\mathvarPluto}{\mathord}{starfontsym}{72}

\graphicspath{{images/}}
\articletype{Original Article}%

\received{21 March 2025}
\revised{18 June 2025}
\accepted{15 July 2025}

\begin{document}

\title{
\texorpdfstring{
\textit{JWST} Imaging of the Closest Globular Clusters -- VI.\\ 
The Lowest-Mass Objects in M\,4 and the Galactic Bulge\protect\thanks{
Based on observations with the  NASA/ESA {\it James Webb Space Telescope},
obtained at the  Space Telescope Science Institute,  which is operated
by AURA, Inc., under NASA contract NAS 5-26555, under GO-1979.
}}
{JWST on M\,4 - Brown Dwarfs}
}

\author[1]{L. R. Bedin*}
\author[2]{R. Gerasimov}
\author[3]{A. Calamida} 
\author[1]{M. Libralato}
\author[1]{M. Scalco}
\author[4,1]{D. Nardiello}
\author[3]{M. Griggio} 
\author[5,6]{D. Apai}
\author[3]{J. Anderson} 
\author[3]{A. Bellini} 
\author[7]{A. Burgasser}

\authormark{\textsc{L. R. Bedin et al.}}

\address[1]{\orgdiv{Istituto Nazionale di Astrofisica}, \orgname{Osservatorio Astronomico di Padova}, \orgaddress{\state{Vicolo dell'Osservatorio 5, Padova, IT-35122}, \country{Italy}}}
\address[2]{\orgdiv{Department of Physics and Astronomy}, \orgname{University of Notre Dame}, \orgaddress{\state{Notre Dame, Indiana, 46556}, \country{USA}}}
\address[3]{\orgname{Space Telescope Science Institute}, \orgaddress{\state{3700 San Martin Drive, Baltimore,  MD 21218}, \country{USA}}}
\address[4]{\orgname{Dipartimento di Fisica e Astronomia "Galileo Galilei" -- Universit\`a degli Studi di Padova}, \orgaddress{\state{Vicolo dell'Osservatorio 3, I-35122 Padova}, \country{Italy}}}
\address[5]{\orgdiv{Department of Astronomy and Steward Observatory}, \orgname{The University of Arizona}, \orgaddress{\state{933 N. Cherry Avenue, Tucson, AZ 85721}, \country{USA}}}
\address[6]{\orgdiv{Lunar and Planetary Laboratory}, \orgname{The University of Arizona}, \orgaddress{\state{1629 E. University Blvd., Tucson, AZ 85721}, \country{USA}}}
\address[7]{\orgdiv{Department of Astronomy \& Astrophysics}, \orgname{University of California San Diego}, \orgaddress{\state{La Jolla, CA 92093}, \country{USA}}}

\corres{*E-mails: luigi.bedin@inaf.it}

\abstract{
We present \textit{James Webb Space Telescope} observations of M\,4 
-- the closest globular cluster -- that probe the lower Main Sequence down to the hydrogen-burning limit. 
The unveiled stellar sequence reaches much fainter luminosities than previously possible, 
revealing a few extremely red 
objects that are consistent with brown dwarfs 
as cool as $T_\mathrm{eff}\sim1000\ \mathrm{K}$.  
However, the lack of a second \textit{JWST} epoch presently prevents us from verifying the cluster membership of these objects. By cross-matching our data with archival \textit{Hubble Space Telescope} images, we are able to verify cluster membership for a subset of objects down to $T_\mathrm{eff}\sim 3000\ \mathrm{K}$. 
The observed color distribution indicate 
that the lower Main Sequence of M\,4 is likely deficient in oxygen compared to its higher-mass post-Main Sequence members by $\sim0.5\ \mathrm{dex}$. This feature has now been observed in three different globular clusters (M\,4, NGC\,6397 and 47\,Tuc), suggesting a general trend. Finally, we derive the mass function of the Galactic bulge in the background of M\,4. The mass function was found to have the bottom-heavy slope of $\alpha=0.88\pm0.36$ and appears to terminate at $\sim 0.15\ \mathrm{M}_\mathSun$, although the latter value may be overestimated due to the limited sample size.
}


\fundingInfo{
This work is based on funding by: 
INAF under the WFAP project, f.o.:1.05.23.05.05;  
The Science and Technology Facilities Council Consolidated Grant ST/V00087X/1; and 
STScI funding associated with GO-1979. 
}

\maketitle

\footnotetext{\textbf{Abbreviations:} 
\textit{JWST}, James Webb Space Telescope; 
\textit{HST}, Hubble Space Telescope; 
$\mathSun$, Sun/Solar
}
\section{Introduction}\label{S:intro}
%
%

Efficient fragmentation of collapsing molecular clouds in the presence of metal cooling \citep{IMF_characteristic_mass_3,IMF_characteristic_mass_4,IMF_characteristic_mass_2,IMF_characteristic_mass_1} and short lifespans of high-mass stars inevitably bias the observed stellar mass function towards cooler and  lower-mass objects. In the solar neighborhood, cool stars and brown dwarfs (BDs) with effective temperatures $T_\mathrm{eff}\lesssim 4000\ \mathrm{K}$ account for over $75\%$ of the stellar census \citep{local_census}. Despite their ubiquity, cooler objects are far more difficult to study and interpret than their warmer counterparts due to both faint magnitudes and increasingly complex physics of low-temperature stellar atmospheres.

The spectra of stars with $T_\mathrm{eff}$ below $\sim3000\ \mathrm{K}$ (so-called \textit{ultracool dwarfs} or UCDs) are dominated by molecular opacity due to $\mathrm{TiO}$, $\mathrm{H_2O}$, $\mathrm{CH_4}$ and other species. Prominent molecular absorption bands span hundreds of nanometers in wavelength and strongly impact not only the spectral energy distribution, but also the underlying atmospheric pressure/temperature structure, and the evolution the star \citep{BC_origin,MIST,2020A&A...637A..38P,roman_omega_cen,2024ApJ...961..139G}. On one hand, this makes the observable properties of UCDs highly sensitive to chemistry, and allows measurements of elemental abundances from low-resolution spectra or broadband photometry \citep{clouds_and_chemistry}. On the other hand, the potential of UCDs as chemical tracers cannot be realized without sophisticated stellar models that account for billions of molecular lines \citep{ExoMol}, and their cumulative impact on the stellar structure and evolution. At $T_\mathrm{eff}\lesssim2500\ \mathrm{K}$, further complexities are introduced by disequilibrium chemistry and condensation that obfuscate the relationship between element abundances and molecular number densities throughout the stellar atmosphere (see reviews in \citealt{cloud_model_comparison,modelling_review}). At present, over a dozen grids of UCD models are available in the literature that attempt to solve the problem of low-temperature stellar modeling with different sets of simplifying approximations (e.g., \texttt{BT-Settl}, \citealt{BT-Settl}; \texttt{LOWZ}, \citealt{LOWZ}; \texttt{Sonora}, \citealt{ElfOwl}; \texttt{ATMO}, \citealt{2020A&A...637A..38P}; \texttt{SAND}, \citealt{SAND}). Unfortunately, the parameters of UCDs inferred from forward modeling of spectra or photometry remain, in many cases, highly dependent on the chosen model grid \citep{AdamTDwarfClassification}.

In order to identify and address the deficits in the current generation of low-$T_\mathrm{eff}$ models, further observational input is required. Of particular value are observations of UCDs, for which some of the properties are known \textit{a priori} from their associations with higher-mass stars. Examples of these so-called \textit{benchmark} UCDs include cool stars and brown dwarfs in binary systems with ``warm'' primaries \citep{BD_binary_benchmarks_1,BD_binary_benchmarks_2,BD_binary_benchmarks_3,BD_binary_benchmarks_4,BD_binary_benchmarks_5}, and UCDs in open star clusters \citep{BD_cluster_benchmarks_1,BD_cluster_benchmarks_2,BD_cluster_benchmarks_3,BD_cluster_benchmarks_4,BD_cluster_benchmarks_5}.

Recently, the \textit{James Webb Space Telescope} (\textit{JWST}) has allowed a systematic study of UCDs in globular clusters (GCs). Prior to \textit{JWST}, large distances to GCs ($\gtrsim5\ \mathrm{kpc}$, \citealt{GC_distances}) imposed a hard faint cutoff of $\sim 0.1\ \mathrm{M}_\mathSun$ ($\sim 3000\ \mathrm{K}$), attainable in the deepest exposures of GCs with the \textit{Hubble Space Telescope} (\textit{HST}) (e.g., \citealt{2001ApJ...560L..75B,2012AJ....143...11K}). 
Below this limit, the mass-luminosity and mass-temperature relationships exhibit abrupt drops to fainter magnitudes and cooler $T_\mathrm{eff}$, as the stellar sequence approaches the hydrogen-burning limit (star / brown dwarf boundary) near $0.08\ \mathrm{M}_\mathSun$ \citep{HBL_3,HBL_4}. A handful of objects in this regime that could be identified in \textit{HST} images did not have sufficiently precise photometry to be differentiable from white dwarfs (WDs) that have similar magnitudes but much bluer colors \citep{2016ApJ...817...48D,BD_hunt}. \textit{JWST} has lifted this limit, allowing GC members as cool as $T_\mathrm{eff}\approx 1400\ \mathrm{K}$ to be identified in images with modest exposure times \citep{paperII}.

Since the majority of GCs are old ($\gtrsim$12\,Gyr; \citealt{GC_ages,vdb13}) and metal-poor ($\mathrm{[Fe/H]}\lesssim\,-1$; \citealt{GC_metallicities}), they may be used to test UCD models in a less explored region of the parameter space that is only sparsely covered by the existing sample of UCD benchmarks due to the strong bias of field objects towards the relatively metal-rich Galactic disk \citep{halo_fraction}. Furthermore, the large masses of GCs \citep{GC_masses} ensure that all evolutionary stages of UCDs are well-represented in the observed color-magnitude diagram (CMD). In particular, GCs are ideally suited to probe the star / brown dwarf transition that is populated by very few UCD candidates (the so-called \textit{stellar/substellar gap}, \citealt{adam_gap,roman_omega_cen,2024ApJ...961..139G}) despite spanning a large range of stellar parameters.

A potential obstacle to using cool GC members as UCD benchmarks is the lack of consensus on whether the chemical abundances in GCs are dependent on stellar mass. Most GCs display a large scatter in light element abundances, which is known as the phenomenon of \textit{multiple populations} (mPOPs) (see the review in \citealt{review_main}). While the origin of mPOPs has not been definitely established, it is suspected that different populations represent distinct generations of star formation \citep{early_NaAl_1}. This scenario appears particularly credible in light of the recent discovery of r-process pollution restricted to only one of the populations in the GC M92 \citep{evan_M92}.

In principle, the association of individual populations in GCs with distinct coeval stellar generations implies that the enrichment of later generations occurred in the gas reservoir of the cluster before star formation took place. In this case, the chemical spread should not directly depend on stellar mass \citep{no_mass_variations_1,2023MNRAS.522.2429M,no_mass_variations_3}. In practice, a difference in the initial mass functions or dynamical histories of individual generations (e.g., see \citealt{2024AN....34540018S}) may still result in over- or under-abundance of stars with enriched chemistry at particular stellar masses. For this reason, it may be challenging to determine if a discrepancy between observed colors of UCD members and low-$T_\mathrm{eff}$ models is caused by a systematic error in the model, or a genuine difference in UCD chemistry. This degeneracy may be addressed by testing the models on multiple GCs with different metallicities and light element abundances, which is one of the main aims of this series of publications.\\ 

The \textit{JWST} program GO-1979 (PI: Bedin) was designed to obtain precise infrared photometry and astrometry of the faintest stars in the two closest Galactic GCs: Messier 4 (M\,4/NGC\,6121, located at 1.85$\pm$0.02\,kpc) and NGC\,6397 (at 2.48$\pm$0.02\,kpc; \citealt{distances}).

The first three papers of the series focused on the WDs (Paper\,I; \citealt{paperI}), BDs (Paper\,II; \citealt{paperII}) and the phenomenon of mPOPs (Paper\,III; \citealt{paperIII}) in NGC\,6397. The fourth entry (Paper\,IV; \citealt{paperIV}) investigated
the chemistry of the lower Main Sequence (MS) in the parallel fields of both clusters.
The most recent publication (Paper\,V; \citealt{paperV}) presented observations of the primary M\,4 field under GO-1979, and delved into the properties of the WD cooling sequence (CS).\\  

This paper (Paper\,VI) revisits the primary field of M\,4 once again, this time focusing on its lower MS down to the stellar/substellar boundary that may provide valuable calibration for state-of-the-art stellar models. Furthermore, the proximity of M\,4 to the Galactic center results in significant contamination of the field by the stellar population of the Galactic bulge. We take this opportunity to carry out a preliminary study of this population as our ancillary science goal.

This paper is organized as follows. In Sect.\,\ref{S:obs}, we briefly summarize the datasets, data reduction and calibration.  In the same section, we also provide a description of the artificial star tests specific to this work, and present the final CMDs. 
In Sect.\,\ref{S:cmd}, we present a comparison of our lower MS observations with theoretical models.  
In Sect.\,\ref{S:bulge}, we focus on the ancillary science case that targets the Galactic bulge. 
In Sect.\,\ref{S:features}, we examine four tentative features in the CMD of M\,4 that warrant follow-up investigations.
Finally, this study is concluded in Sect.\,\ref{S:Conclusions}.

%
\section{Observations, data-reduction, calibrations, Artificial Star Tests and the Color-Magnitude Diagram}\label{S:obs}
\label{Sec:DATA}
%
The observations used in this work were first presented in Paper\,V. 
The relevant
\textit{JWST}
images were captured on April 9, 2023, under program GO-1979, using a 6-point FULLBOX dither pattern, with simultaneous 515.365\,s exposures in NIRCam SW (F150W2) and LW (F322W2) ultra-wide filters. The field is centered at (RA,Dec)$=$($245^\circ.9297$, $-26^\circ.4504$) and spans 1.9-8 arcmin from the cluster center. The images were processed using the \textit{JWST} pipeline, converting pixel values to counts and flagging unusable data via DQ flags. Enhanced dynamic range was achieved by incorporating \texttt{frame\,zero} for saturated pixels.

We also employ observations, obtained with \textit{Hubble Space Telescope} (\textit{HST}) at an earlier epoch, to estimate astrometric proper motions. The \textit{HST} images used in this work originate from program GO-10146 (L. Bedin, 2004) and were obtained with the Advanced Camera for Surveys (ACS) in the Wide Field Channel (WFC). In 2004 (July-August, $\sim$2004.607), 20 exposures ($\sim$1200\,s each) used the F606W filter. In June 2005, four additional $\sim$1200\,s exposures were taken with the ACS/WFC F775W filter.
The combination of \textit{HST} images from $\sim20$ years ago and new \textit{JWST} data allowed us to estimate proper motions for well-measured sources in the overlap region between the two surveys, which corresponds to approximately 1/20 of the full NIRCam field of view (FoV, see Paper\,V). Unlike Paper\,V, the photometric analysis carried out in this study is based exclusively on \textit{JWST} magnitudes.

The data reduction procedures for the \textit{JWST} dataset were described in Paper\,I, while 
reduced data was presented in detail in Paper\,V. Here, we restrict our discussion of the process to a brief summary. 
NIRCam images were reduced using the methods detailed in the first three papers of the \textit{"Photometry and Astrometry with JWST"} series \citep{2022MNRAS.517..484N,2023AN....34430006G,2023MNRAS.525.2585N}. This approach, 
previously applied also to the GC 47\,Tuc \citep{2023MNRAS.521L..39N,2025A&A...694A..68S}, 
uses a two-pass photometry method as defined by \citet{2008AJ....135.2055A}.
The first-pass photometry measures source positions and fluxes via effective-PSF fitting, correcting for geometric distortion as in \citet{2023AN....34430006G}. Results are aligned to a common reference frame based on Gaia DR3, enabling consistent comparison.
Second-pass photometry refines detections and measurements of faint sources using a modified version of the code 
by \citet{2008AJ....135.2055A} for ACS/WFC, named \texttt{KS2} and presented in \citet{2017ApJ...842....6B}. 
\texttt{KS2} outputs include fluxes, positions, and quality diagnostics, 
such as RMS brightness errors, PSF fit quality, and a stellarity index (\texttt{RADXS}; \citealt{2008ApJ...678.1279B}).
The photometric calibration was to the Vega-magnitude system following the procedures outlined by \citet{2023MNRAS.521L..39N} and \citet{2005MNRAS.357.1038B}. 
In the following, for the calibrated magnitudes we will adopt the symbols $m_{\rm F150W2}$ and $m_{\rm F322W2}$.\\

\begin{figure*}
\centerline{\includegraphics[width=168mm]{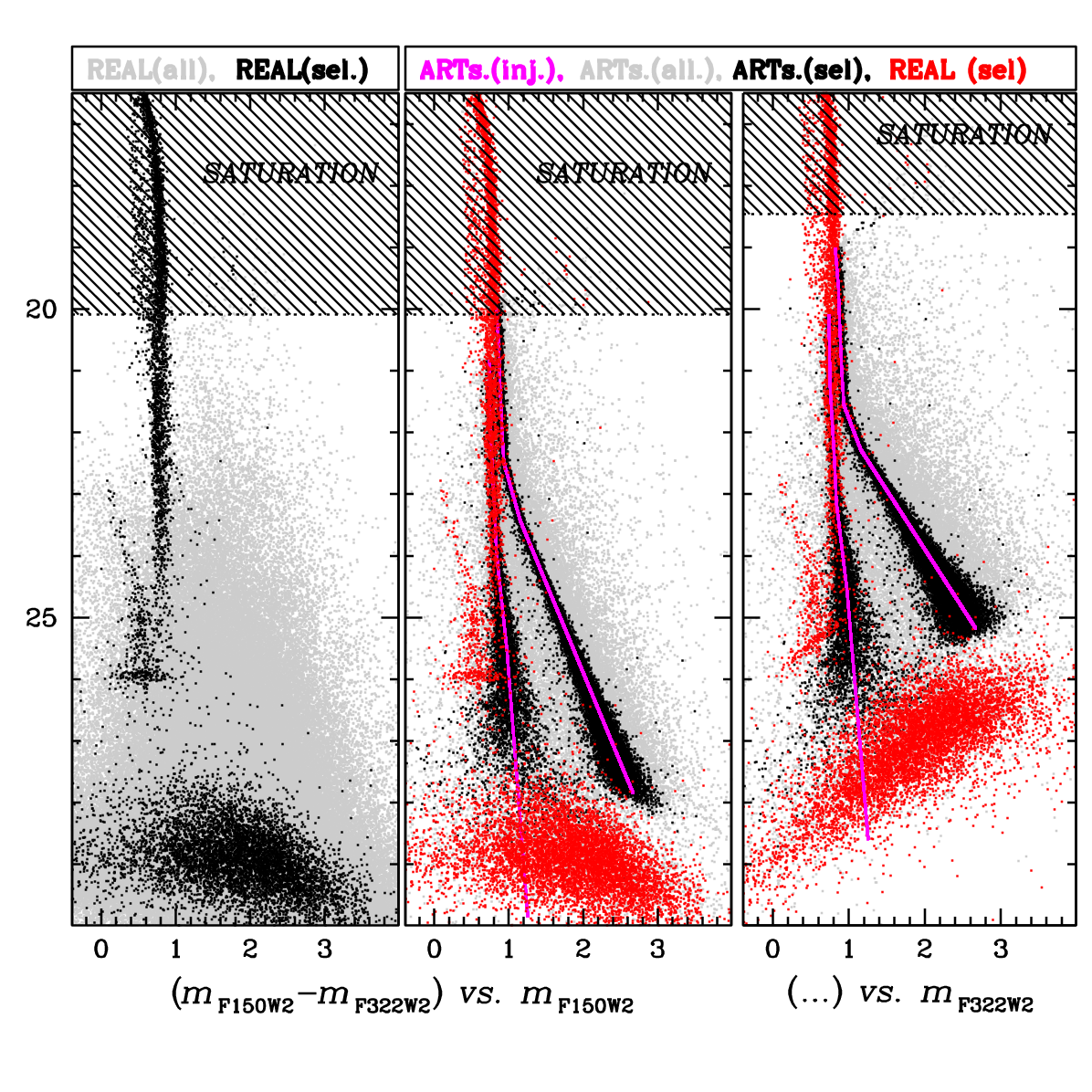}}
\caption{
\textit{Left panel} shows the 
$(m_{\rm F150W2}-m_{\rm F322W2})\,vs.\, m_{\rm F150W2}$ CMD for \textit{all} \textit{`real'}
local maxima detected within the NIRCam FoV (grey points). 
Among those, `real' sources that passed all the photometric quality tests 
are marked in black (selected, i.e., \textit{sel.}). 
The bulk of objects at color around 2 and magnitudes  $m_{\rm F150W2}\sim 29$ indicate 
the floor-noise level. 
The onset of saturation is highlighted by a shaded area for sources brighter 
than $m_{\rm F150W2}\sim 20$ (and labelled).  
\textit{Middle panel} shows the same CMD, this time for the artificial stars (ARTs.), 
with injected (i.e., \textit{`inj.', in} magenta) and recovered (grey/black) magnitudes of artificial stars. 
For reference, well-measured observed `real' sources are overplotted
in red (same as the black points in the left panel). 
\textit{Right panel} shows the analogous CMD but with $m_{\rm F322W2}$ on the y-axis. 
Note that here saturation happens at $m_{\rm F322W2}\sim 18.5$.  
}
\label{fig:ASTs}
\end{figure*}
%

A key difference in our data reduction compared to Paper\,V lies in the placement of the artificial star tests (ASTs). 
Instead of distributing them along the white dwarf cooling sequence, we placed them along two arbitrarily defined sequences 
in the CMDs. These sequences were chosen to encompass the regions where the majority of low-mass and very low-mass stellar objects are expected --both cluster members and Galactic field stars.
We performed ASTs along two fiducial sequences defined as follows. The first was manually drawn to trace the apparent sequence of M4 stars down to $m_{\rm F150W2} \simeq 22$, and extended to fainter magnitudes along the envelope of the reddest stars in the CMDs, as determined by eye. The second sequence was selected to follow the main distribution of non-member stars, which are expected to primarily belong to the Galactic bulge, given the proximity of the field to the Galactic center
(see Sect.\,\ref{S:bulge}).
For each fiducial sequence we generated 100,000 artificial stars with randomly chosen positions within the NIRCam FoV and random F150W2 magnitudes, drawn from the uniform distribution between $m_{\rm F150W2}=20$ and $\sim$28. The color of each star was set to follow the adopted fiducial lines. 
The stars were injected one at a time to avoid overcrowding and measured in the same way as real sources. They were considered recovered when the measured positions and magnitudes fell within 1\,pixel and $-2.5 \log{2}$\,($\simeq0.75$)\,magnitudes of the true values, respectively.

In the left panel of Fig.\,\ref{fig:ASTs} we present the CMD in 
$(m_{\rm F150W2}-m_{\rm F322W2})\,vs.\, m_{\rm F150W2}$ space
for all the detected real local maxima (grey points) within the NIRCam FoV. 
The sources that passed our photometric selection (i.e., well-measured real point sources, 
see Papers\,I and V) are highlighted with black points. 
The saturation region of the CMD with magnitudes brighter than $m_{\rm F150W2}\sim 20$ and $m_{\rm F322W2}\sim18.5$ 
is shaded in the figure.
The other two panels focus on artificial star tests 
shown with the same code (gray for all, and black for those passing the selections), 
but also show ---for reference--- the  selected real sources in red.
In the middle panel, the same CMD for recovered artificial stars
In magenta are shown the injected magnitudes. 
Finally, in the right panel of Fig.\,\ref{fig:ASTs} we show the analogous CMD, but in the
$(m_{\rm F150W2}-m_{\rm F322W2})\,vs.\, m_{\rm F322W2}$ observational plane, i.e., placing the redder filter on the y-axis. 

We release the photometric catalogue in electronic format (both in the journal 
as supplementary files and at our webpage\footnote{
\texttt{https://web.oapd.inaf.it/bedin/files/PAPERs\_eMATERIALs/\-JWST/GO-1979/P06/}. 
}). 
For artificial stars we also provided the inserted and recovered magnitudes. 
For both real and artificial stars we provide a flag that indicate which 
sources have passed the quality selections. We also provide the sample of 
proper-motion selected members.

%
%
\section{Comparison with Models}\label{S:cmd}
%
%
In Paper\,II of this series, we introduced \texttt{SANDee} -- a new grid of model isochrones based on the \texttt{MESA} evolutionary code \citep{MESA,MESA_2,MESA_3,MESA_4,MESA_5}, and \texttt{SAND} model atmospheres \citep{SAND}. At present, the \texttt{SANDee} grid is the only set of theoretical isochrones that spans the temperature range of our data and the metallicity of M\,4. 
In this section,  we compare the observed CMD of confirmed cluster members to the nearest \texttt{SANDee} isochrone. 
Our comparison with models for the low-mass MS stars is limited to the $(18.5 \leq m_{\rm F322W2} \leq 21)$ segment of the CMD, as it reliably passes the photometric cut and is not saturated. 
This segment corresponds to stars with masses between 
$\sim0.25\ \mathrm{M}_\mathSun$ and $0.1\ \mathrm{M}_\mathSun$, 
and effective temperatures 
$3700\ \mathrm{K}\gtrsim T_\mathrm{eff}\gtrsim 3000\ \mathrm{K}$, 
i.e. the segment includes some of the coolest stars in the cluster, leading into the UCD regime ($T_\mathrm{eff}\lesssim 3000\ \mathrm{K}$). Since the hypothetical mass dependence of the chemical spread in the cluster is expected to be more pronounced at lower masses (i.e., at stellar masses that are most removed from the spectroscopically accessible upper MS and post-MS members), our sample establishes a lower limit on the extent of this phenomenon at $T_\mathrm{eff}\ll 3000\ \mathrm{K}$.

\texttt{SANDee} isochrones are parameterized by the overall metallicity, $[\mathrm{Fe/H}]$, and the enhancement of $\alpha$-elements, $[\mathrm{\alpha/Fe}]$, where $\alpha$-elements are $\mathrm{O}$, $\mathrm{Ne}$, $\mathrm{Mg}$, $\mathrm{Si}$, $\mathrm{S}$, $\mathrm{Ar}$, $\mathrm{Ca}$ and $\mathrm{Ti}$. Of these elements, $\mathrm{O}$ has the largest effect on the observed $(m_{\rm F150W2}-m_{\rm F322W2})$ color of low-mass members due to the proximity of the \qty{3}{\micro\meter} $\mathrm{H_2O}$ absorption feature to the F322W2 band. 
In general, lower $\mathrm{[O/Fe]}$ increases flux in the F322W2 band and makes the color of the star redder. By contrast, the effect of other $\alpha$-elements on the $(m_{\rm F150W2}-m_{\rm F322W2})\,vs.\, m_{\rm F322W2}$ CMD is far more subtle (see Paper\,IV).

In this study, we are primarily interested in the potential of low-mass GC members as UCD benchmarks, whose abundances may be independently inferred from the spectroscopic observations of bright post-MS stars, under the assumption that the statistical distribution of abundances is independent of stellar mass. Based on the spectroscopic abundance analysis in \citet{2008A&A...490..625M}, the post-MS members of M\,4 have the average metallicity of $[\mathrm{Fe/H}]=-1.07$ and the average oxygen abundance of $[\mathrm{O/Fe}]=0.4$ with $1$-sigma scatter of $\sim 0.1\ \mathrm{dex}$. The $[\mathrm{Fe/H}]=-1.1$, $[\mathrm{\alpha/Fe}]=0.35$ \texttt{SANDee} isochrone is closest to these values, so we used it in our analysis. Given the sample size in \citet{2008A&A...490..625M} ($93$ measurements) and the offset in $[\mathrm{O/H}]$ between the spectroscopic average and the closest \texttt{SANDee} model ($0.08\ \mathrm{dex}$), the theoretical isochrone 
should be bluer than $21\pm 4\%$\footnote{This percentage was estimated as the expected fraction of members with $[\mathrm{O/H}]$ values lower than that of the isochrone (i.e. lower than $-1.1+0.35=-0.75$) in a sample of $93$ values drawn from a normally distributed population with the mean and standard deviation found in \citet{2008A&A...490..625M} ($-1.07+0.4=-0.67$ and $0.1$, respectively).} of the UCD members assuming that the effect of small corrections in color due to elements other than oxygen is small, and ignoring all other systematic effects.

\begin{figure}
\includegraphics[width=\columnwidth]{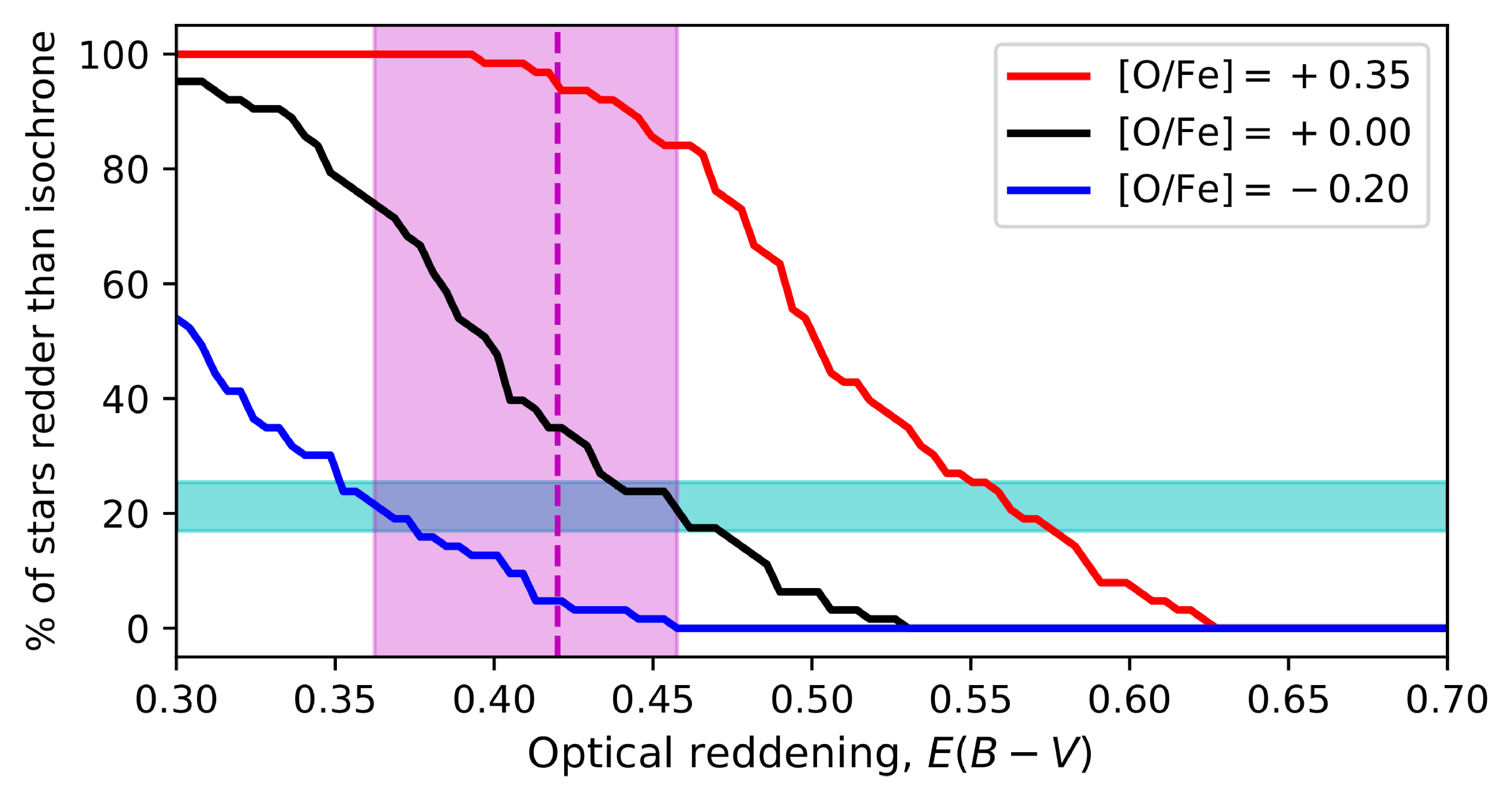} 
\caption{Comparison of the spectroscopically inferred distribution of $[\mathrm{O/Fe}]$ in M\,4 and the photometric spread of the lower MS. As detailed in text, the \texttt{SANDee} isochrone with $[\mathrm{Fe/H}]=-1.1$ and $[\mathrm{O/Fe}]=[\mathrm{\alpha/Fe}]=0.35$ is expected to be bluer than $21\pm4\%$ (cyan shading) of the stars in the magnitude range $(18.5 \leq m_{\rm F322W2} \leq 21)$, assuming that the distribution of $[\mathrm{O/Fe}]$ is the same across the CMD. The red curve shows the actual fraction of stars redder than the isochrone as a function of $E(B-V)$. The black and blue curves provide the same statistic for alternative isochrones with reduced $[\mathrm{O/Fe}]$. The magenta shading indicates the expected range of $E(B-V)$ near the line of sight of our observations. The vertical magenta line highlights the expected $E(B-V)$ at the center of the field.}
\label{fig:frac_bluer}
\end{figure}

Due to the apparent proximity of M\,4 to the $\rho$\,Oph cloud complex, our line of sight is highly extinguished, and the reddening law does not obey the standard total-to-selective extinction ratio of $R_V=3.1$. For this reason, we cannot adopt a fixed reddening value, $E(B-V)$, in the model, and must treat it as a free parameter. This parameter plays a key role in determining the actual number of stars that are redder than the theoretical isochrone. In Fig.\,\ref{fig:frac_bluer}, we plotted the fraction of confirmed cluster members (see Sec.\,\ref{S:bulge}) redder than the closest \texttt{SANDee} isochrone as a function of $E(B-V)$ (red curve). We also indicated the reddening at the center of the observed field, $E(B-V)=0.42$, and the range of $E(B-V)$ within a $15$ arcmin radius, $0.36\rightarrow 0.46$, taken from the reddening map in \citet{reddening_map}. We used \texttt{BasicATLAS} \citep{BasicATLAS} to transform the \texttt{SANDee} isochrone to the observed plane, adopting the distance of $1.85\ \mathrm{kpc}$ \citep{distances}, the age of $12\ \mathrm{Gyr}$ \citep{2009ApJ...697..965B}, $R_V=3.6$ \citep{2012HendricksM4red} and the reddening law from \citet{2023ApJ...950...86G}.

The specific choice of the reddening law (\citealt{2023ApJ...950...86G} used by us vs. \citealt{CC89} used in \citealt{2012HendricksM4red}) or the statistical error in the adopted value of $R_V$ ($\pm0.07$) do not noticeably impact our results, as the expected difference in synthetic color due to these effects does not exceed $0.01\ \mathrm{mag}$ for $T_\mathrm{eff}=3000\ \mathrm{K}$.

Fig.\,\ref{fig:frac_bluer} demonstrates that $E(B-V)\sim0.56$ is required to reconcile the spectroscopic abundances with the lower-MS photometry, which is well above the estimated range of $E(B-V)$ near the line of sight of our observations. The corresponding difference in synthetic color at $T_\mathrm{eff}=3000\ \mathrm{K}$ calculated for $E(B-V)=0.56$ and $E(B-V)=0.42$ (reddening at the center of the field from \citealt{reddening_map}) is $\approx0.08\ \mathrm{mag}$, which is much larger than the effect of the adopted reddening law ($\sim0.01\ \mathrm{mag}$) or the uncertainty in photometric zero points ($\sim0.02\ \mathrm{mag}$\footnote{\href{https://jwst-docs.stsci.edu/jwst-calibration-status/nircam-calibration-status/nircam-imaging-calibration-status}{https://jwst-docs.stsci.edu/jwst-calibration-status/nircam-calibration-status/nircam-imaging-calibration-status}}).
This indicates either a discrepancy between the chemical composition of lower-MS and post-MS stars, or a systematic error in the adopted model. 
To quantify this discrepancy, we also plotted in Fig.\,\ref{fig:frac_bluer} the expected number of confirmed members redder than the isochrone for two modified versions of the closest \texttt{SANDee} isochrone, which we calculated in Paper~IV. These alternative isochrones have $[\mathrm{O/Fe}]=0$ (black curve) and $[\mathrm{O/Fe}]=-0.2$ (blue curve). They predict a good match between spectroscopic and photometric $[\mathrm{O/Fe}]$ at the edges of the acceptable $E(B-V)$ range. Therefore, the oxygen abundance on the lower MS of M\,4 appears to be between $0.4\ \mathrm{dex}$ and $0.6\ \mathrm{dex}$ lower, in the absence of systematic errors. This offset is much larger than the difference between the solar $\mathrm{[O/H]}$ estimated in \citet{2008A&A...490..625M} and the standard solar $\mathrm{[O/H]}$ in \citet{solar_oxygen} ($\approx 0.07\ \mathrm{dex}$). Therefore, the observed discrepancy in $[\mathrm{O/Fe}]$ cannot be explained by inaccurate solar calibration.

\begin{figure*}
\includegraphics[height=84mm]{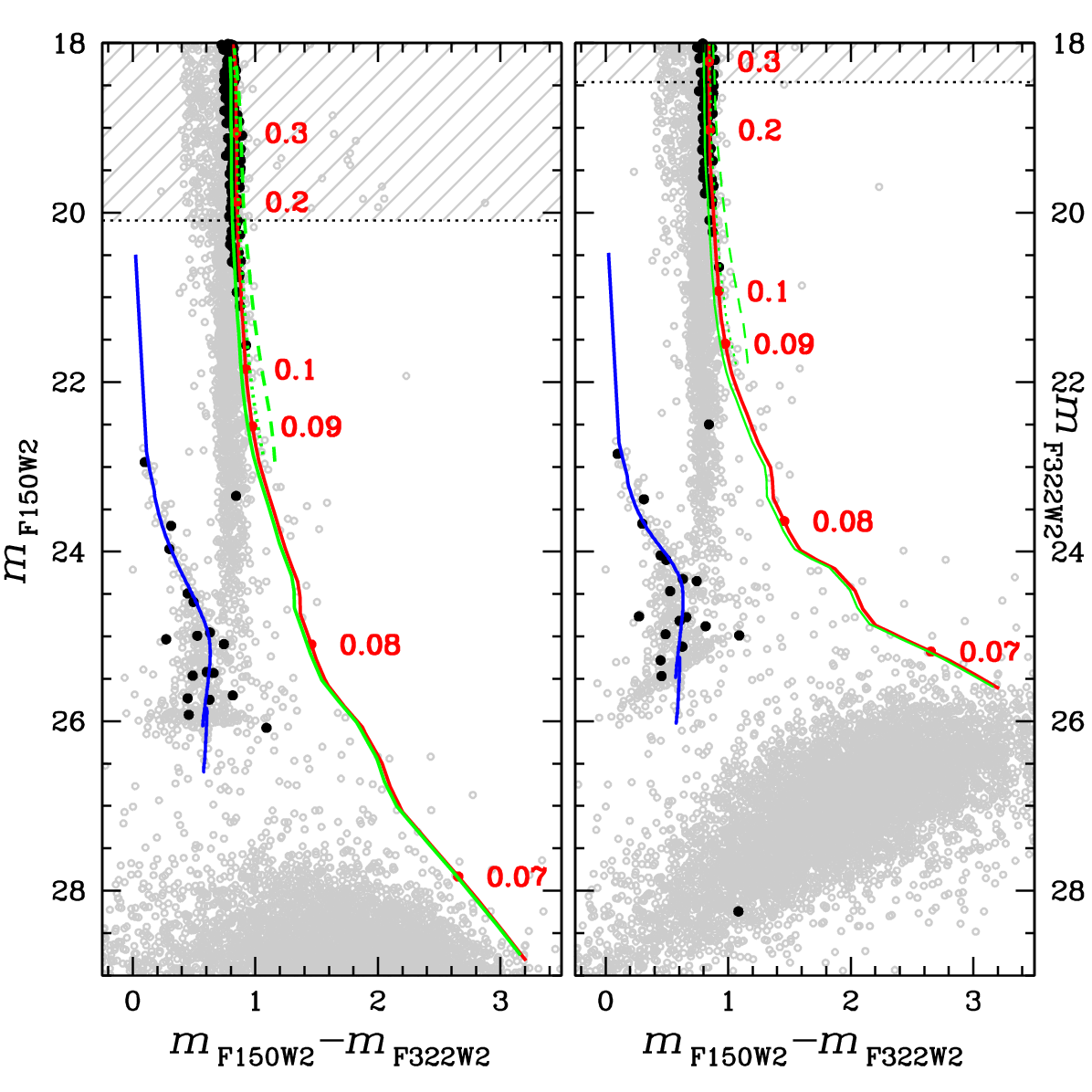} 
\includegraphics[height=84mm]{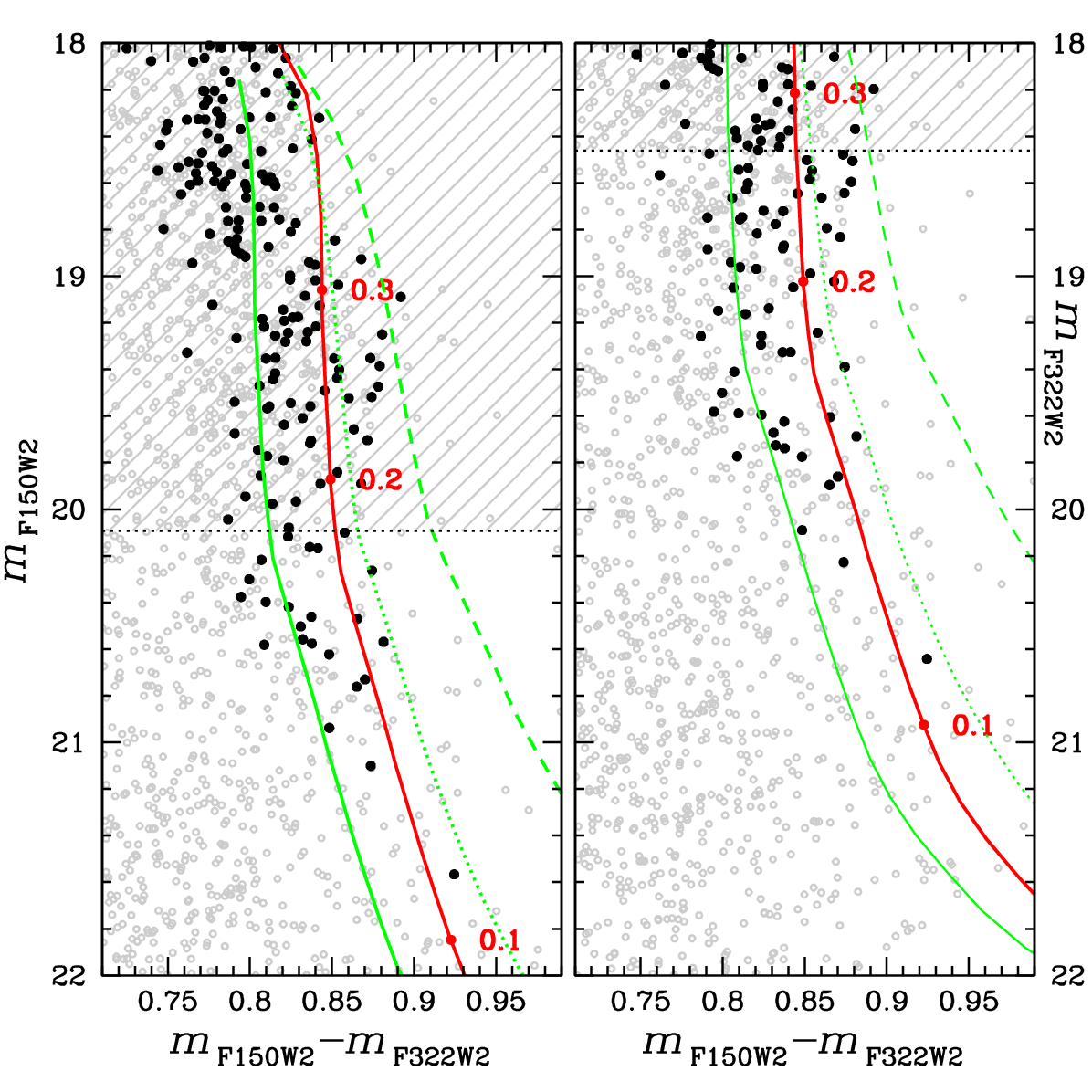} 
\caption{
\textit{(Left Panels:)} 
Comparison of observations with isochrones, for both MS and WD CS. 
The two CMDs were obtained from NIRCam@\textit{JWST} images collected under program 
GO-1979. 
Saturation limit is shaded as in Fig.\,\ref{fig:ASTs}. Grey circles are sources that passed our photometric selection, 
while black points are a subset of sources with verified cluster membership based on proper motion (see text). 
The 12-Gyr MS+BD isochrone (red lines) is a \texttt{SANDee} model (Paper\,II) with the closest chemical composition to the spectroscopic measurements in \citet{2008A&A...490..625M} and the reddening value of $E(B-V)=0.5$. This reddening value was chosen to approximate the median color-magnitude trend of M\,4 on the lower MS. The solid green lines are based on the same isochrone but at a more realistic reddening of $E(B-V)=0.42$, adopted from \citet{reddening_map}. The dotted and dashed green lines are the modified versions of this isochrone with $[\mathrm{O/Fe}]=0.0$ and $[\mathrm{O/Fe}]=-0.2$, as described in text. 
Selected stellar masses are labelled along the red isochrone in $M_\mathSun$. For completeness, we also show the 12-Gyr WD isochrone presented in Paper\,V (blue lines). 
\textit{(Right Panels:)} same plots but zoomed in at the lower MS for clarity. 
}
\label{fig:ISO}
\end{figure*}

All three isochrones used in our analysis are plotted in Fig.\,\ref{fig:ISO} for $E(B-V)=0.42$ \citep{reddening_map}. As seen in Fig.\,\ref{fig:frac_bluer}, the $[\mathrm{Fe/H}]=-1.1$, $[\mathrm{\alpha/Fe}]=0.35$ \texttt{SANDee} isochrone serves as the approximate median of the observed photometric scatter at the reddening of $E(B-V)\approx 0.5$. This configuration was adopted in the rest of our analysis, and is also plotted in Fig.\,\ref{fig:ISO}.\\

%
%

%
%

\begin{figure*}
\centerline{\includegraphics[width=168mm]{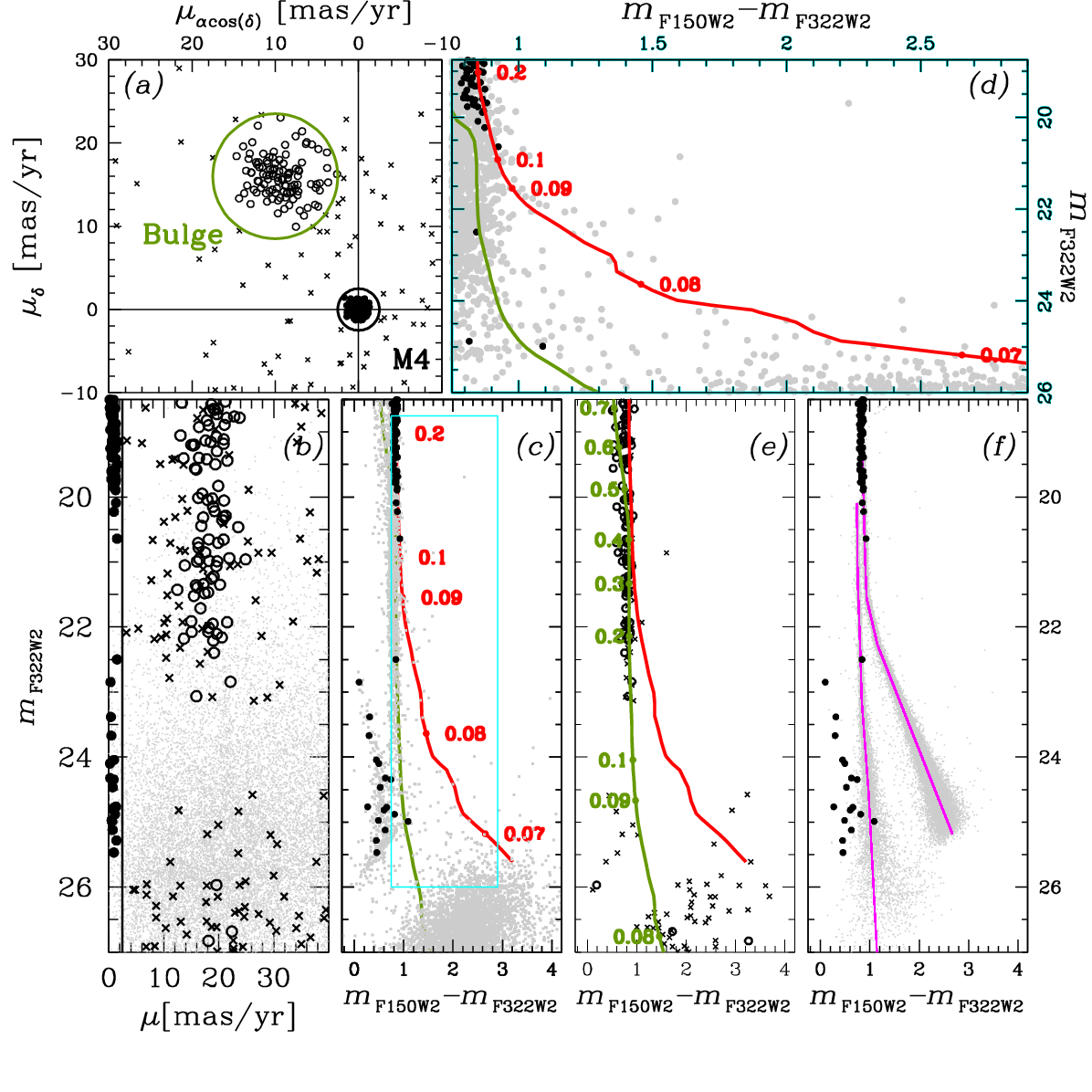}}
\caption{
\textit{(a)} The VPD of sources in the overlap region between the \textit{HST} and \textit{JWST} datasets that have passed our photometric selection. The origin (marked as the intersection of the vertical and horizontal lines) 
was chosen to coincide with the average motion of M\,4. The green and black circles delineate our membership selection for the Galactic bulge and M\,4, respectively.  
M\,4 members are shown with filled circles, Galactic bulge members with small open circles (defined as falling within 2.5\,$\sigma$ of the bulk motion of the bulge given by \citealt{2003AJ....126..247B}), non-members of both are shown with crosses.
\textit{(b)} Combined proper motion vs.\ $m_{\rm F322W2}$. Same symbols as in \textit{(a)}. The sources with proper motion measurements that failed our photometric selection are shown in grey. 
\textit{(c)} $(m_{\rm F322W2}-m_{\rm F322W2})$ vs.\, $m_{\rm F322W2}$ CMD for all sources 
that passed the photometric selection within the NIRCam FoV (in gray) and for the sources that also satisfied the M\,4 membership criterion in the overlap region (filled circles). 
The red isochrone is the \texttt{SANDee} model introduced in Sect.\,\ref{S:cmd}; reference masses are labeled in $\mathrm{M}_\mathSun$. 
The green curve is the same isochrone, but shifted to the average distance of the bulge members in our NIRCam field. A cyan rectangle shows a portion of the CMD, blown-up in panel 
\textit{(d)}. This zoom-in of the CMD highlights a small group of stars with no proper motion (and therefore no membership information) 
that are more-or-less aligned with the theoretical isochrone for M\,4 (red). 
Panels \textit{(e)} and \textit{(f)} show the same CMD as in panel \textit{(c)} but for field objects (bulge and disk members) and artificial stars (Sect.\,\ref{S:obs}), respectively. Proper motion-verified members of M\,4 are highlighted with filled circles in the latter.
}
\label{fig:VPD}
\end{figure*}
%

%
\section{The Outskirts of the Bulge}\label{S:bulge}
%
%
The line of sight of our M\,4 field at the galactic coordinates $(\ell,\mathscr{b}) \simeq (-9^\circ\!.03,+15^\circ\!.97)$ is situated $\sim$18$^\circ\!.35$ from the center of the Milky Way, and therefore probes the outskirts of the Galactic bulge, which has a projected corotation radius of $(17\pm3)^\circ$ \citep{corotation_radius}.

As the saying goes, \textit{"One man's noise is another man’s signal"}. In this section, we explore our ancillary science case by probing the low-mass MS stars of the Galactic bulge, which may be used as tracers of its stellar component. This brief study also emphasizes the value of our astro-photometric catalogs for studying the Galactic field at large.  

In panel $(a)$ of Fig.\,\ref{fig:VPD}, we show the vector-point diagram (VPD) for the sources with calculated proper motions, 
i.e., those within the overlap region between \textit{HST} and \textit{JWST} observations that passed the photometric selection. 
Here M\,4 members are indicated with filled circles (with proper motions within 2.5\,mas\,yr$^{-1}$ of the cluster average), while 
Galactic bulge members are shown with open circles ("$\bigcirc$", defined to fall within 7.5\,mas\,yr$^{-1}$ of the Galactic bulge over-density, which we took to be at $(\mu_{\alpha\cos{\delta}};\mu_{\delta})\simeq(10;16)$\,mas\,yr$^{-1}$, following \citealt{2003AJ....126..247B}). 
The remaining sources (mainly members of the Galactic disk) or low-quality measurements are indicated with crosses ("$\times$"). 
The background extra-galactic extended objects without a point-source profile were largely filtered out by the \texttt{RADXS} selection (see Paper\,I and references therein).
Note that the internal proper motion dispersion of M\,4 members is expected to be approximately 0.5\,mas\,yr$^{-1}$
\citep{GC_distances}, 
which is much smaller than the 2.5\,mas\,yr$^{-1}$ threshold 
adopted for our membership criterion (thick black circle centered at 0,0). This generous selection was set to accommodate proper motion uncertainties. 
The proper motion dispersion of the Galactic bulge is expected to fall below 3\,mas\,yr$^{-1}$ \citep{2003AJ....126..247B} ($\sim$100\,km\,s$^{-1}$ at $\sim$\,8\,kpc). Therefore the adopted limit of 7.5\,mas\,yr$^{-1}$ for the Galactic bulge membership (green circle) corresponds to a rather tight 2.5\,$\sigma$ threshold, which should remove a considerable fraction of disk stars.

In panel $(b)$, we show the one-dimensional combined proper motions, defined as 
$\mu=\sqrt{\mu_{\alpha\cos{\delta}}^2+\mu_{\delta}^2}$,  
as a function of the $m_{\rm F322W2}$ magnitude. 
For reference, the panel also shows the proper motions of the sources that did not pass the photometric selection in grey.   

In panel $(c)$, we show the $(m_{\rm F150W2}-m_{\rm F322W2})\,vs.\, m_{\rm F322W2}$ CMD for 
M\,4 member stars that passed both photometric and membership selections (black points), and sources across the full NIRCam field that only passed the photometric selection (grey points) but have no proper motions. 
We also overplotted the \texttt{SANDee} isochrone described in Sec.\,\ref{S:cmd} for the MS/BDs (red curve).
Additionally, we show the same isochrone in green, 
shifted to match the distance to the bulk of the Galactic bulge stars in our field.
The outer region of the Galactic bulge that we are probing is expected to be at an 
average distance of $\mathcal{R} \sim 7.8$\,kpc\footnote{
$\mathcal{R} = \mathcal{R}_\circ \cos{\ell} \cos{\mathscr{b}}$ \citep{2003MmSAI..74..436B, 2003AJ....126..247B}; 
assuming $\mathcal{R}_\circ = 8.2$\,kpc \citep{2019A&A...625L..10G}. 
}. 
Concerning the extinction, 
once we reach the distance of M\,4 --approximately 1.85\,kpc from us and located at a 
Galactic latitude of $\sim$16$^\circ$--  we are already more than 500\,pc above the Galactic plane, where reddening is expected to be negligible. It is therefore reasonable and widely accepted to assume negligible reddening beyond this height.
[For completeness, most of the reddening along the line of sight to M\,4 arises from the $\rho$\,Ophiuchi cloud complex, situated at a distance of about 200\,pc in the Scorpius–Ophiuchus region 
\citep{2012HendricksM4red}, and thus well in the foreground of M4.]
Although the Galactic bulge stars in our field have an intrinsic spread in distance, age, and chemical composition, 
it is noteworthy that the isochrones computed for M\,4, once shifted to the expected distance to the Galactic bulge, trace the overall distribution of the Galactic bulge MS well.
Interestingly, some of the outliers among the MS of M\,4 approximately coincide with the bulge-shifted (green) isochrone, suggesting that they may actually be members of the Galactic bulge with poorly estimated proper motions.

\begin{figure}
\includegraphics[height=84mm]{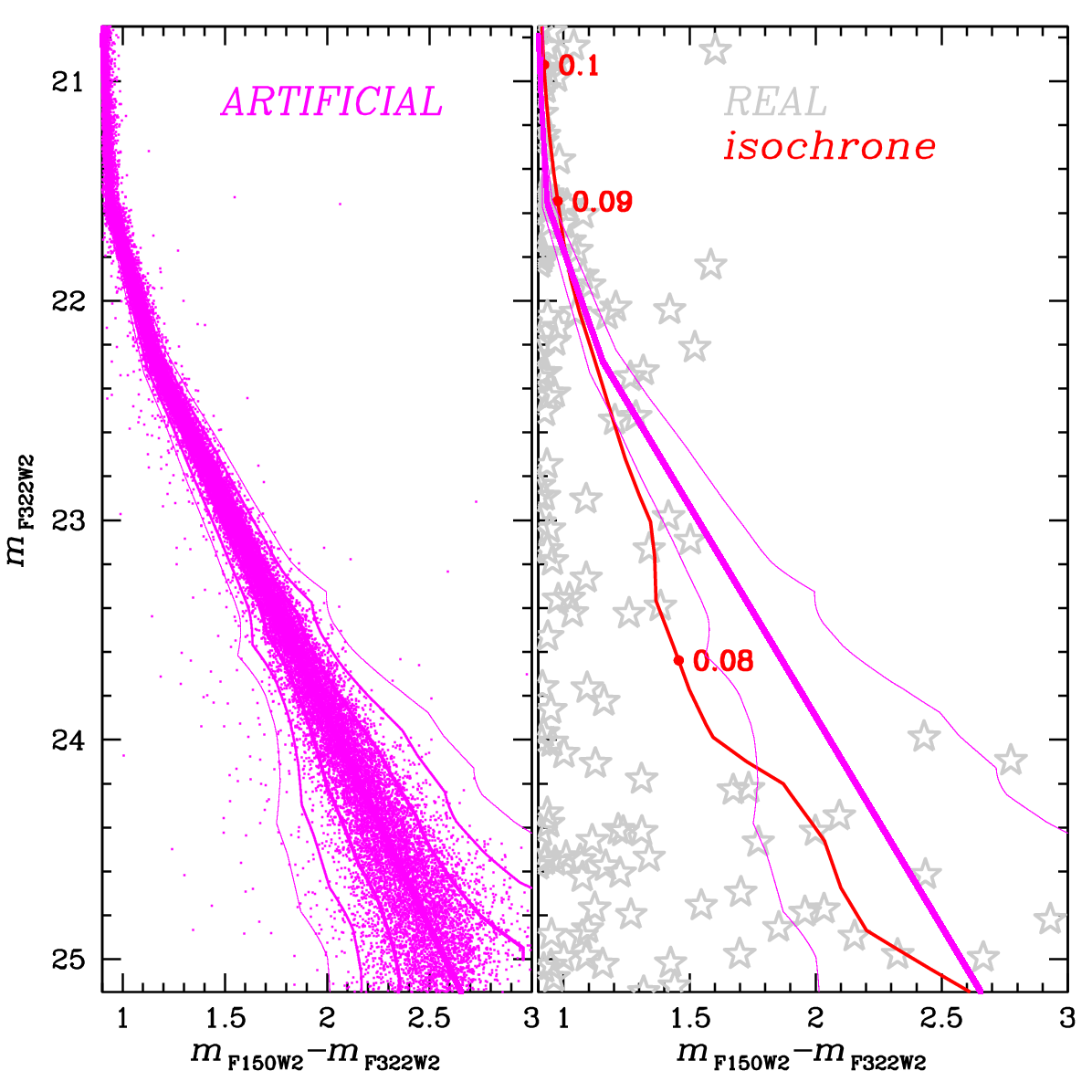} 
\caption{
A detail of the CMD in the region of interest of the brightest BDs, with the aim of comparing 
ASTs (on the left), `real' sources on the right, with both ischrones and the estimated photometric errors. 
\textit{Left:} panel shows the ASTs along the reddest fiducial sequence, as they were recovered (dots) and how 
they were injected (thick magenta line). The thin lines mark the estimated 
1-2-3\,$\sigma$ of photometric errors around the fiducial.  
\textit{Right:} The fiducial and 3-$\sigma$ lines are compared with real stars (star symbols) and
the isochrone employed in previous figures. 
Given the large uncertainties in both observations and models, many of the sources are compatible with 
being BDs members of M\,4, but need astrometric follow-up.  
}
\label{fig:ITEM9}
\end{figure}

Panel $(d)$ shows a zoom-in of panel $(c)$, confined to the region outlined with the cyan box. This panel reveals an apparent sequence of sources without proper motion measurements (and therefore indicated with grey points) that are situated close to the M\,4 isochrone, and are therefore likely to be members of the cluster with $T_\mathrm{eff}<3000\ \mathrm{K}$. 
This tentative extension of the MS appears to reach at least $m_{\rm F322W2} \sim 24$, and possibly fainter. 
In Fig.\,\ref{fig:ITEM9} we better compare the CMD for the ASTs and real stars 
in the same regions where the brightest BDs could be detected.
For comparison, the hydrogen-burning limit predicted by \texttt{SANDee} (Paper\,II) at the adopted chemical composition of M\,4 falls at $\sim0.08\ \mathrm{M}_\mathSun$, also labeled in the panel. Therefore, some of these sources may be BD members of M\,4, although a proper motion confirmation of membership is required before any definitvie conclusions can be drawn.

%
%

%
In panel $(e)$ of Fig.\,\ref{fig:VPD}, we show the same CMD as in panel $(c)$, but for field objects, including both Galactic bulge members (marked with "$\bigcirc$") and candidate disk stars (along with objects with poorly measured proper motions, marked with "$\times$"). In panel $(f)$, the same CMD is shown for ASTs, with injected and recovered magnitudes in magenta and gray, respectively. 

A comparison of panels $(e)$ and $(f)$ clearly indicates that our observations have reached a genuine drop in the luminosity function (LF) of the Galactic bulge MS, which occurs near  $m_{\rm F322W2}\sim23$, where the sample completeness remains high. 

In the case of the Galactic bulge, $m_{\rm F322W2}\sim23$ mag corresponds to a mass of 
$\sim 0.15\ \mathrm{M}_\mathSun$.
\citet[][hereinafter, CA15]{calamida2015} present the bulge mass function (MF) 
in the SWEEPS (Sagittarius Window Eclipsing Extrasolar Planet Search) low-reddening window, reaching 
a mass of $\sim 0.15\ \mathrm{M}_\mathSun$ with  $\approx$ 50\% completeness, 
while \citep{zoccali2003} reached a similar limit in the Baade's window. The 
lack of stars fainter than $m_{\rm F322W2}\sim23$\,mag in this part of the Galactic bulge could be due to small number statistics: only 83 stars were selected as members of the Galactic bulge, some of which may have been confused with residual contamination from the Galactic disk.

\begin{figure}
\centerline{\includegraphics[width=0.5\textwidth]{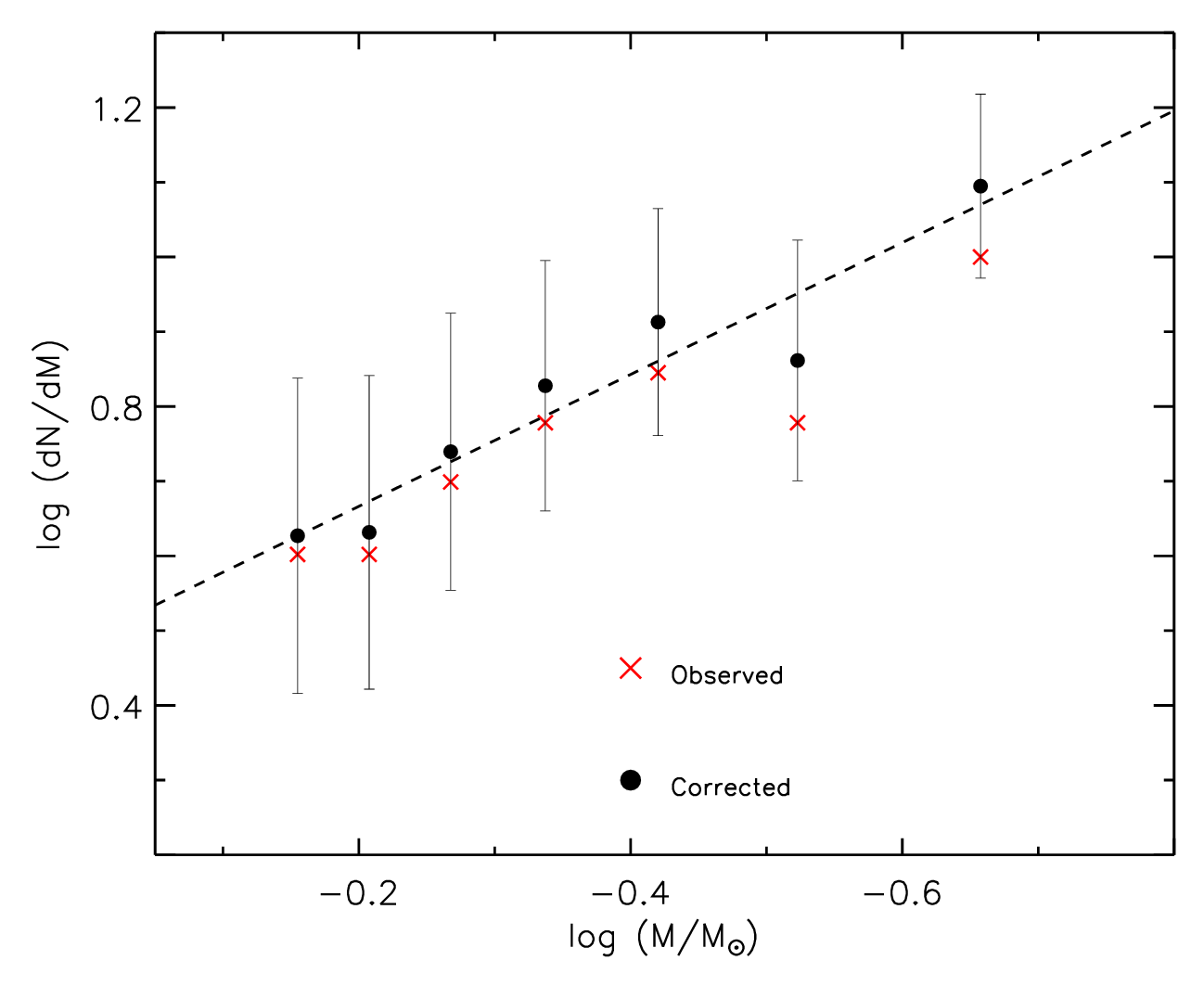}}
\caption{Observed (red cross) and completeness-corrected (filled circles) MF in the Galactic bulge based on 45 stars.
The dashed line is the best fit to the observed MF, i.e., a single-component power-law with slope $\alpha = 0.88\pm0.36$ (see text). Error bars are shown.}
\label{fig:bulge_imf}
\end{figure}

A detailed analysis of the stellar component of the Galactic bulge is beyond the scope 
of this study. Nonetheless, this preliminary analysis demonstrates that the astro-photometric catalogs obtainable with NIRCam on \textit{JWST} can open a new window of opportunities 
for studying the mixture of populations in the bulge of the Milky Way.

As a test, we used candidate Galactic bulge stars to construct a MF 
along the observed line of sight. To convert magnitudes into masses, we used a 
BASTI\footnote{http://basti-iac.oa-abruzzo.inaf.it/tracks.html} 
scaled-solar isochrone for an age of 
11\,Gyr and solar metallicity, [Fe/H]\,=\,0.06  \citep[$Z = 0.0172$]{pietrinferni2004}; 
a more appropriate isochrone to represent the bulk of the Bulge's population 
than the one employed in Sect.\,\ref{S:cmd}.
Note that the MS mass-luminosity relation is largely insensitive to age for ages older 
than $\sim$ 2\,Gyr, and weakly dependent on metallicity (CA15). We used the distance of $\sim$ 7.8\,kpc, and the reddening 
value of $E(B-V) =$ 0.37\,mag \citep{paperV}. 

In this analysis, we only considered objects fainter than $m_{\rm F322W2}\sim 18.75$\,mag to avoid 
saturation effects, resulting in the sample of 45 stars. Fig.~\ref{fig:bulge_imf} shows the observed (red crosses) and completeness-corrected MFs (black 
filled circles) for the selected Galactic bulge members. The completeness ranges from 96\% 
at the bright end to $\sim$ 70\% at $m_{\rm F322W2} \sim$ 23\,mag. The MF extends 
from $\sim0.7$ to $\sim 0.2\ \mathrm{M}_\mathSun$. 
Errors based on Poisson statistics are shown in the figure and, as expected, are quite large. We proceeded to fit the the completeness-corrected MF (black filled circles in Fig.~\ref{fig:bulge_imf}) with a 
single-component power law model ---\,$\xi(M)=dN/dM\propto M^{-\alpha}$\,---
and obtained a slope of $\alpha =$\,0.88$\pm$0.36 (dashed line). This value is in good agreement with the single-component 
power law slope derived by CA15 for the Galactic bulge MF in the SWEEPS window, 
$\alpha =$ 1.14$\pm$0.10, assuming the absence of binaries. Unfortunately, our sample size does not allow to adequately account for the effect of unresolved binaries on the MF. 

On the other hand, the IMF for the SWEEPS field from CA15 presents a shallower slope of 0.89 in the low-mass regime ($M \lesssim 0.5 M_{\odot}$ when the presence of no binaries is assumed).
We do not have enough statistics to fit our MF with two components as well as simulate the binary contamination; in addition, the bright part of the MF might be affected by stars being close to the saturation limit. However, the fit of our MF is weighted more towards low masses, demonstrating the good agreement of this MF with the SWEEPS one. 

We further evaluated the effect of reddening and adopted distance modulus on the inferred MF. We found that increased or reduced $E(B-V)$ along the line of sight by $\pm$ 0.1\,mag 
changes the mass estimates by $\lesssim$ 2\%. A similar variation in the distance modulus 
by $\pm$ 0.1\,mag ($\pm$ 400 Kpc) offsets the mass estimates by $\lesssim$ 8\%. Similar findings  were reported by CA15 for the mass estimates of Galactic bulge stars in the SWEEPS field.

Due to the small FoV and sample size, the MF inferred in this work must be considered with caution. In a future study, we intend to obtain proper motion-verified membership for a larger number of stars, in order to enable a more comprehensive analysis.

\section{Candidate Features in the CMD of M\,4 awaiting confirmation}
\label{S:features}
In this section we bring the readers' attention to four putative features that 
we attempt to highlight
in the CMD of point sources within our M\,4 \textit{JWST} NIRCam FoV. 
These sources lack proper motion measurements, and therefore their membership 
must be verified in a future study. 
Nonetheless, these features may correspond to genuine components of the stellar population of M\,4 in various evolutionary stages.

\begin{figure*}
\centerline{
\includegraphics[height=89mm]{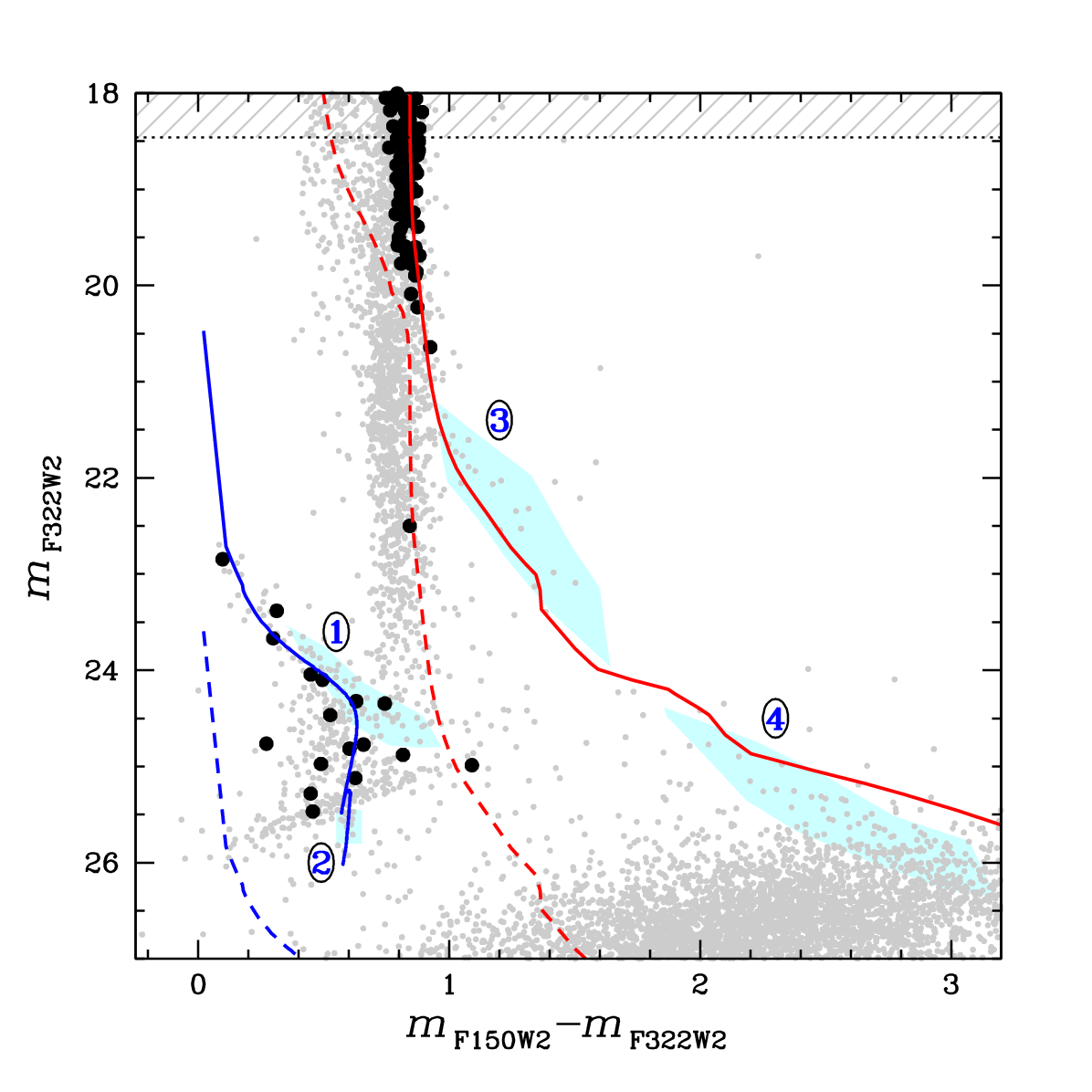}
\includegraphics[height=89mm]{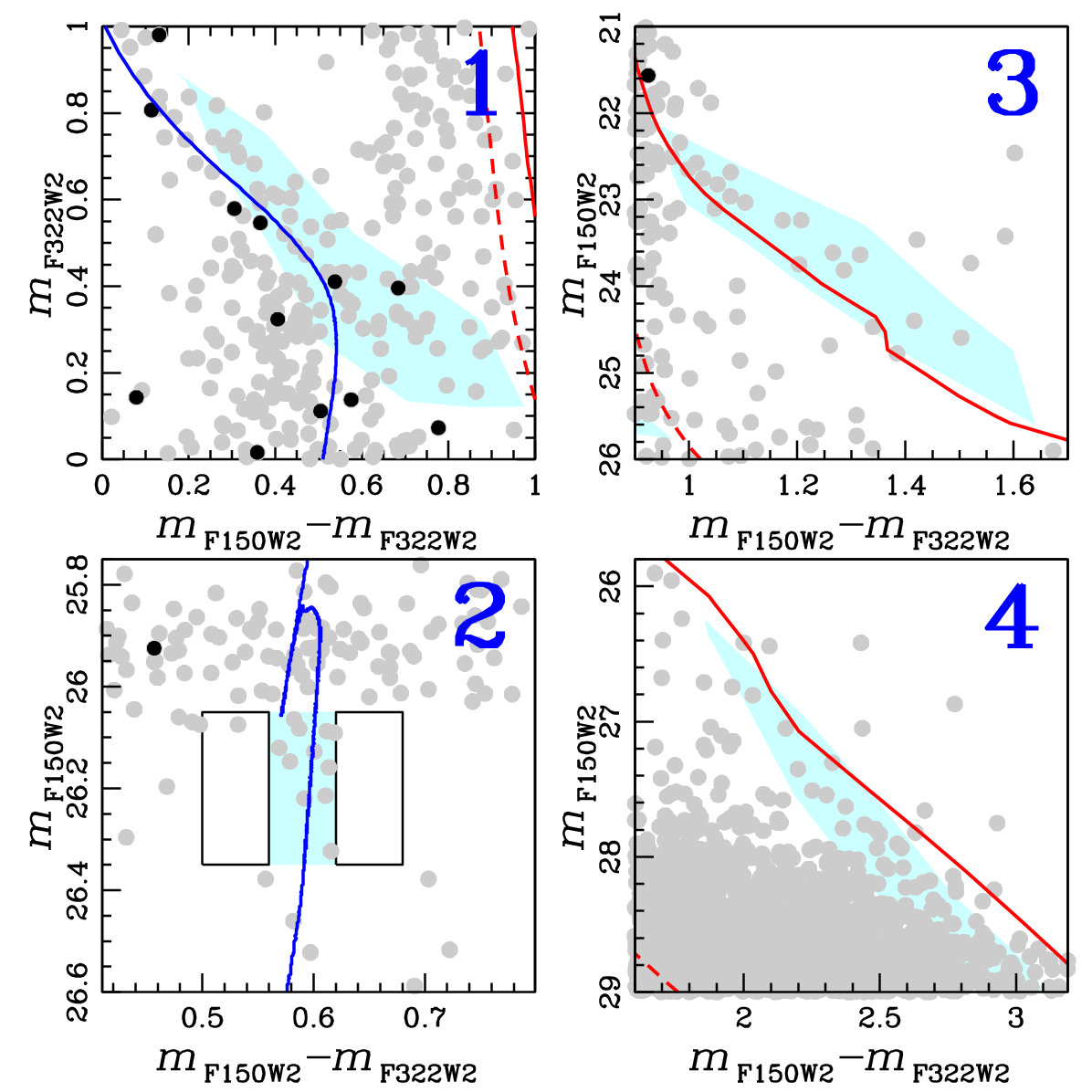}
}
\caption{
\textit{(Left:)}
In this CMD we highlight with cyan regions four potentially real features which deserve future investigations (see text). 
Isochrones for M\,4 as in Fig.\,\ref{fig:ISO} (and shifted at Bulge's distance with dashed lines). 
\textit{(Right:)}
Zoomed-in views around each of the four putative features, with their identification numbers labeled in the top-right corner of each panel.
(\#1): 
The cyan-shaded region highlights a possible broadening of the WD\,CS toward redder colors --somewhat reminiscent of the feature seen in NGC\,6397 (see Paper\,I).
(\#2): To illustrate the potential significance of this feature, we count the number of sources within three equal-sized rectangular regions in the CMD. 
%
(\#3): The cyan region contains sources that approximately follow the expected location of very low-mass stars in M\,4, as predicted by the red isochrone.
(\#4): This is the most uncertain—but also the most intriguing—feature, as it may represent a population of BDs belonging to M\,4. Only follow-up observations—with improved photometric depth and proper motion data—will be able to reduce contamination from spurious detections and confirm cluster membership. [See text for discussion].
}
\label{fig:FEATs}
\end{figure*}

The features of interest are shaded in cyan in the CMD of sources with no proper motions (gray points) presented in Fig.\,\ref{fig:FEATs}, 
and marked with labels from (1) to (4). 
The four panels on the right provide zoom-in view of each one. 
Our tentative interpretation of these features follows.

\begin{itemize}
    \item[(1)] The first feature is a putative sequence of sources just above the WD\,CS. 
    Note that, as previously discussed in Paper\,V, the WD\,CS appears significantly broader 
    than expected based on estimated photometric errors. This highlights a possible broadening 
    of the WD CS toward redder colors --somewhat reminiscent of the feature seen in NGC\,6397 
    (see Paper I). 
    These could be WDs members with infrared  excess due to circumstellar disks \citep{2021MNRAS.504.2707G},  
    but if confirmed as members, detailed comparison with models must be carried out in a follow-up study to test this hypothesis \citep{paperI}; 
    \item[(2)] the second lies just below the peak of the WD CS LF 
    (at $m_{\rm F322W2}\simeq$25.5 and $m_{\rm F150W2}\simeq$26.2), and 
    could be related to the most massive WDs in the cluster; 
    To illustrate the potential significance of this feature, in the left-panel 
    labelled with "2", we count the number of sources within three equal-sized 
    rectangular regions in the CMD. The width of these boxes is approximately equal to the expected photometric color errors (see Paper\,V).
    The central (cyan) box, which contains the feature, includes 11 sources, while the two adjacent boxes (blueward and redward) contain only 1 and 0 (respectively). 
    While not conclusive, this asymmetry is suggestive of a real feature under simple 
    Poisson statistics.
    \item[(3)] the third is most likely the faintest part of the MS, tracing UCDs down to the hydrogen-burning limit, which is expected around $m_\mathrm{F322W2}\approx23.2$ (and $m_\mathrm{F150W2}\approx24.6$) according to the adopted \texttt{SANDee} model;
    symmetric regions above and below this sequence do not show comparable numbers of objects.
    \item[(4)] and finally, the most uncertain feature: a putative sequence of objects that are situated close to the model-predicted BD sequence of the cluster. 
    This is the most uncertain feature, but also the most intriguing ---potentially involving 
    BDs belonging to M\,4.
\end{itemize}

To asses the legitimacy of these features, we require a second \textit{JWST} NIRCam epoch, which would not only improve the overall photometry of this dataset, but also, most importantly, provide reliable proper motion-verified cluster membership over the entire NIRCam FoV where it is unaffected by crowding. 
Furthermore, proper motions for the entire NIRCam FoV would increase the sample size of stars with confirmed membership by a factor of $\sim$20.

A second \textit{JWST} epoch would be most valuable to assess the nature of extremely red objects (features (3) and (4)); as these sources are completely inaccessible with \textit{HST}, and no membership information is available even in the small portion of the FoV that has both \textit{JWST} and archival \textit{HST} images. 

%
\section{Conclusion}
\label{S:Conclusions}

This work marks \textit{JWST}’s first attempt to explore the faintest MS stars in M\,4 -- the nearest Galactic GC. The new observations are likely deep enough to probe the transition between hydrogen-burning stars and BDs; however, the presently available sample is limited by the lack of a second \textit{JWST} epoch and, for this reason, does not reveal any objects less massive than 0.1\,M$_\mathSun$ with confirmed cluster membership. 
Many sources, though lacking proper motion measurements, appear consistent with predictions for BDs down to 0.07\,M$_\mathSun$ and $T_\mathrm{eff}\sim 1000\ \mathrm{K}$. 

The limited number of stars with proper motion-verified membership -- only about 1/20 of the FoV, concentrated in the most crowded region -- hampers a detailed quantitative analysis of the LF. Such analysis (for both MS/BDs and WDs) will require a second epoch of observations to establish cluster membership across the entire NIRCam FoV observed in this first \textit{JWST} epoch.

In the meantime, we focus on insights provided by the positions of the MS stars in the CMD.
The key findings are summarized below. 

\begin{itemize}
    \item In Sect.\,\ref{S:cmd}, we showed that the colors of lower-MS members of M\,4 are inconsistent with the spread in the oxygen abundance inferred from the spectroscopy of bright post-MS members of the cluster. This discrepancy suggests that at least one of the following statements is true:
    \begin{enumerate}
        \item The interstellar reddening along the observed line of sight is much higher than reported in the literature ($E(B-V)\gtrsim0.55$, as opposed to the upper bound of $E(B-V)<0.46$ in \citealt{reddening_map}), or the extinction law is anomalously steep in the infrared.
        \item The \texttt{SAND} model atmospheres that form the basis for the \texttt{SANDee} isochrone adopted in this study suffer from major systematic errors that are equivalent to an underestimation of the strength of $\mathrm{H_2O}$ absorption in the F322W2 band by over $50\%$ at $T_\mathrm{eff}$ as high as $3500\ \mathrm{K}$, i.e., the temperature regime far warmer than the typical onset of complex low-temperature effects in UCD atmospheres.
        \item The scatter in the oxygen abundance on the lower MS of M\,4 is offset by $0.4-0.6\ \mathrm{dex}$ compared to the higher-mass post-MS members.
    \end{enumerate}
    We consider the third possibility most likely, since a similar discrepancy has been identified in the GC NGC\,6397 (see Paper\,IV) and 47\,Tuc \citep{2025A&A...694A..68S}. The latter study is especially significant, as it identified an ambiguous discontinuity in the lower MS of 47\,Tuc (named the ``kink'' in \citealt{2025A&A...694A..68S}, also observed by \citealt{2024ApJ...965..189M}). This feature is most readily associated with the early onset of $\mathrm{CH_4}$ absorption, which requires a large oxygen deficit in the atmosphere. M\,4 is now the third GC where such deficit has been identified. We therefore conclude that UCDs in GCs may not be suitable as benchmarks for low-$T_\mathrm{eff}$ model atmospheres.
    \item As demonstrative ancillary science, we presented a pilot application of \textit{JWST} to the study of the stellar component of the Galactic bulge. While a detailed analysis of this population is beyond the scope 
    of this paper, we used the current NIRCam@\textit{JWST} astro-photometric catalog to derive the bulge MF along the observed line of sight. We also showed that the MF is well approximated by a single-component power law model with $\alpha=0.88\pm0.36$, in good agreement with previous results for other parts of the Galactic bulge based on \textit{HST} data. This analysis demonstrates how the non-member contaminants in \textit{JWST} and \textit{HST} fields near the Galactic center can be used to probe the properties of the Galactic bulge at different longitudes and latitudes. Note that the \textit{Roman Space Telescope} 
    is set to begin operations in 2027 and it will observe most of the Galactic 
    plane and numerous bulge fields in different near infrared filters. The completeness and 
    depth will vary with position, but for some regions, astrometry (also in 
    combination with previous observations) and photometry will be precise (and 
    deep) enough to study and compare the disk and bulge MFs.
    
    Our investigation demonstrates how \textit{JWST} photometry can unveil the low-mass tail of the Galactic bulge MF. We expect our preliminary results to be vastly improved by a second epoch of \textit{JWST} observations, which would provide 
    proper motion measurements for a statistically significant sample of stars.
    \item Finally, we identify four potential features in the CMD of M\,4 that warrant further investigation. These features include the extended morphology of the WD\,CS that includes both over- and under-luminous WDs compared to the model prediction, a tentative extension of the UCD sequence in the cluster across the hydrogen-burning limit, and a possible presence of a few
    extremely faint and red objects that are associated with the anticipated BD sequence of M\,4.
\end{itemize}

%
%


%
\section{Acknowledgments}
\noindent
We are grateful to the anonymous Referee for their constructive feedback, 
which significantly contributed to improving the manuscript.
We warmly thank STScI, our Program Coordinator and Instruments Reviewers 
--Shelly Meyett, Mario Gennaro, Paul Goudfrooij and David Golimowski-- 
for their great support during the review of our problematic observations.
LRB, ML, MS, DN, and MG acknowledge support by INAF under the WFAP project, f.o.:1.05.23.05.05. 
%
RG, DA, JA, and ABe, acknowledge support from STScI funding associated with GO-1979. 
%


\newpage
~\\ 


\bibliography{biblio}

\end{document}